\documentclass[aps,prd,twocolumn,superscriptaddress,notitlepage,nofootinbib,preprintnumbers]{revtex4-1}
\usepackage[utf8]{inputenc}
\usepackage{CJKutf8}
\usepackage{epsfig}
\usepackage{multirow}
\usepackage{slashed}
\usepackage{amsmath}
\usepackage{amssymb}
\usepackage{physics}
\usepackage{qcircuit}
\usepackage{braket}
\usepackage{graphicx}
\usepackage{subfigure}
\usepackage{color}
\usepackage{epstopdf}
\usepackage[inline]{enumitem}

\def\bea#1\eea{\begin{align}#1\end{align}}

\newcommand{\bef}{\begin{figure}[!htp]}
\newcommand{\eef}{\end{figure}}

\usepackage{xcolor}

\usepackage[colorlinks,
linkcolor = blue,
anchorcolor=blue,
citecolor=blue,
urlcolor=blue]{hyperref}

\begin{document}
\preprint{RIKEN-iTHEMS-Report-26}
\title{Jianmu-Q: Bridging Quantum Field Theory and Quantum Hardware with AI Agents}

\date{\today}

\author{Xiaolin Wang}
\email{xlwangn@m.scnu.edu.cn}
\affiliation{State Key Laboratory of Nuclear Physics and Technology, Institute of Quantum Matter, South China Normal University, Guangzhou 510006, China}
\affiliation{Key Laboratory of Atomic and Subatomic Structure and Quantum Control (MOE),
Guangdong-Hong Kong Joint Laboratory of Quantum Matter, Guangzhou 510006, China}

\author{Xingyu Guo}
\email{guoxy@m.scnu.edu.cn}
\affiliation{State Key Laboratory of Nuclear Physics and Technology, Institute of Quantum Matter, South China Normal University, Guangzhou 510006, China}
\affiliation{Guangdong Basic Research Center of Excellence for Structure and Fundamental Interactions of Matter, Guangdong Provincial Key Laboratory of Nuclear Science, Guangzhou 510006, China}

\author{Jinyang Li}
\email{jinyang.li@riken.jp}
\affiliation{RIKEN Center for Interdisciplinary Theoretical and Mathematical Sciences (iTHEMS), Wako, Saitama 351-0198, Japan}
\affiliation{KEK Theory Center, Institute of Particle and Nuclear Studies, KEK, Tsukuba, Ibaraki 305-0801, Japan}
\affiliation{Graduate University for Advanced Studies (SOKENDAI), Oho 1-1, Tsukuba, Ibaraki 305-0801, Japan}

\author{Tianyin Li}
\email{tianyin.li@riken.jp}
\affiliation{RIKEN Center for Interdisciplinary Theoretical and Mathematical Sciences (iTHEMS), Wako, Saitama 351-0198, Japan}

\author{Enke Wang}
\email{wangek@scnu.edu.cn}
\affiliation{State Key Laboratory of Nuclear Physics and Technology, Institute of Quantum Matter, South China Normal University, Guangzhou 510006, China}
\affiliation{Guangdong Basic Research Center of Excellence for Structure and Fundamental Interactions of Matter, Guangdong Provincial Key Laboratory of Nuclear Science, Guangzhou 510006, China}

\author{Lingxiao Wang}
\email{lingxiao.wang@riken.jp}
\affiliation{RIKEN Center for Interdisciplinary Theoretical and Mathematical Sciences (iTHEMS), Wako, Saitama 351-0198, Japan}
\affiliation{Institute for Physics of Intelligence, Graduate School of Science, The University of Tokyo, Bunkyo-ku, Tokyo 113-0033, Japan}

\author{Hongxi Xing}
\email{hxing@m.scnu.edu.cn}
\affiliation{State Key Laboratory of Nuclear Physics and Technology, Institute of Quantum Matter, South China Normal University, Guangzhou 510006, China}
\affiliation{Guangdong Basic Research Center of Excellence for Structure and Fundamental Interactions of Matter, Guangdong Provincial Key Laboratory of Nuclear Science, Guangzhou 510006, China}
\affiliation{Southern Center for Nuclear-Science Theory (SCNT), Institute of Modern Physics, Chinese Academy of Sciences, Huizhou 516000, China}

\date{\today}        

\begin{abstract}
Quantum computing offers a promising approach to non-perturbative simulations of quantum field theories, particularly for real-time dynamics and systems affected by the sign problem.
Yet translating continuum field theories into quantum computations remains a challenging interdisciplinary task.
Here we introduce \texttt{Jianmu-Q}, a multi-agent framework that transforms natural-language descriptions of quantum field theories into quantum simulation workflows.
Specialized agents handle theoretical formulation, lattice discretization, qubit mapping, and circuit design, while physical consistency checks and independent numerical calculations provide validation throughout the workflow.
We benchmark the framework on three field theories of increasing complexity: the $(1+1)$-dimensional Nambu--Jona-Lasinio and Schwinger models, and a $(2+1)$-dimensional $\mathbb{Z}_2$ gauge theory. Across these systems, noiseless quantum-circuit simulations converge toward independent exact-diagonalization and tensor-network results, while selected circuits are executed on IBM quantum hardware, demonstrating an end-to-end, validation-guided approach to quantum simulation of field theories.
\end{abstract}

\maketitle

\section{Introduction}
\label{sec:introduction}
Many crucial problems in quantum chromodynamics (QCD) are intrinsically non-perturbative, such as confinement, hadron structure, and the QCD phase diagram. Lattice QCD provides a first-principles framework for such calculations~\cite{Wilson1974Confinement,Hagler:2009ni,Lin:2017snn,Briceno:2017max,FlavourLatticeAveragingGroupFLAG:2024oxs}. However, traditional Monte Carlo lattice QCD in Euclidean space suffers from the notorious sign problem for real-time dynamics and at finite baryon density~\cite{Troyer:2004ge,Aarts2016FiniteDensity,Nagata:2021ugx,Aarts:2026uiu}: its computational cost grows exponentially with the spacetime volume, so current finite-density results largely rely on extrapolations from zero or imaginary chemical potential~\cite{deForcrand:2002hgr,Bazavov:2017dus,Borsanyi:2020fev}.
This motivates the search for a sign-problem-free, non-perturbative method to study quantum field theory (QFT), including QCD.

\begin{figure*}[t]
    \centering
    \includegraphics[width=1\textwidth]{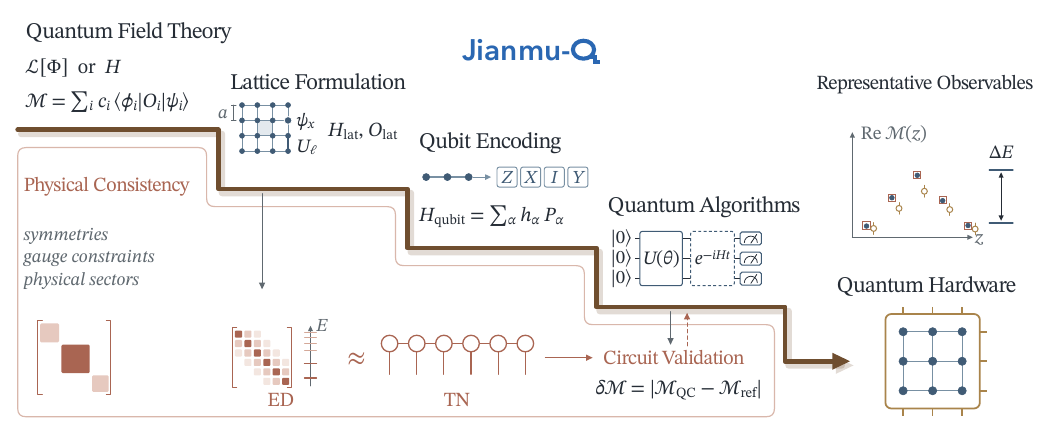}
    \caption{\texttt{Jianmu-Q} as a staircase from quantum field theory to quantum hardware. Specialized agents carry field-theory specifications: a Lagrangian or a lattice Hamiltonian, external states, and an observable $\mathcal{M}$ [Eq.~\eqref{eq:physical_matrix_element}], through a regularized lattice formulation, a model-specific qubit encoding, and task-dependent circuits (state preparation, optional dynamics [dashed], measurement) to quantum hardware, the intended simulation platform, subject to present device feasibility. Beneath the steps, independent validation comprises physical-consistency checks, ED and TN references for the same regulated finite-lattice problem, and a comparison of noiseless circuit results with them, including convergence in the ansatz depth $D$ and, where applicable, the Trotter step count $N_T$; failed checks trigger targeted refinement (dashed). The schematic readout shows $\mathrm{Re}\,\mathcal{M}(z)$ and $\Delta E$ from ED/TN (squares), noiseless circuits (dots), and hardware (open circles); validation addresses the finite lattice, not the continuum limit.}
    \label{fig:workflow}
\end{figure*}

As a natural alternative, quantum computation in Minkowski space offers a direct way to represent quantum states and simulate their evolution without a sign problem~\cite{Feynman1982Simulation,Lloyd1996Simulation}. The feasibility of simulating QFT on quantum computers has been established, together with algorithms for scattering in scalar field theory~\cite{Jordan2012QFT}. For gauge theories, experimental and quantum algorithm studies of the Schwinger model have been reported~\cite{Martinez2016Dynamics,Kokail2019Verification,Klco:2018kyo,Shaw:2020udc,Chakraborty:2020uhf,Honda:2021aum,Sakamoto:2023cxs,Gustafson:2024bww,Pomarico:2025ngp}, and there are also many works simulating non-Abelian and higher-dimensional lattice gauge theories (LGT)~\cite{Alexandru:2019nsa,Paulson:2020zjd,Yamamoto:2020eqi,Davoudi:2022xmb,Ciavarella:2024fzw,Halimeh:2024bth,Crippa:2024hso,Hanada:2025goy,Hayata:2026rmv,Li:2026ppp}. Quantum algorithms have also been proposed for simulating various quantities and phenomena in QCD, such as hadron structure~\cite{Mueller:2019qqj,Lamm2020Parton,Kreshchuk:2020aiq,Li:2021kcs,Li:2022lyt,Ikeda:2025bjb,Zou:2026cfk,Li:2026thq}, scattering amplitudes~\cite{Jordan2012QFT,Li:2023kex,Bennewitz:2025nhz}, hadronization and jet physics~\cite{Bauer:2019qxa,Florio:2023dke,Li:2024nod,De:2024smi}, and the QCD phase diagram~\cite{Zhang:2024fgv,Xie:2022jgj,Czajka:2021yll,Yamamoto:2021vxp}. Together, these developments form a broader program to apply quantum simulation to high-energy physics~\cite{Banuls:2019bmf,Bauer:2022hpo,Funcke:2023jbq,DiMeglio:2023nsa}.

However, turning these physical problems into an executable quantum circuit requires expertise across different research fields. A continuum model must first be discretized into a finite lattice theory with specified boundary conditions and physical sectors~\cite{Smit2002Lattice,Gattringer2010Lattice} before its Hamiltonian and observables are encoded on qubits~\cite{Jordan1928Pauli,Shaw:2020udc}. Then state preparation and measurement procedures have to preserve the same conventions~\cite{Peruzzo2014VQE,Pedernales:2014izf}, and the designed circuits must be tested and adapted to the available computational resources. The output of the quantum circuit also needs to be cross-checked against independent classical calculations. Each step is familiar within its own field, but their combination requires expertise in QFT, Hamiltonian LGT~\cite{KogutSusskind1975}, quantum algorithms, and numerical methods such as exact diagonalization (ED) and tensor networks (TN)~\cite{White1992DMRG}. In practice, such interdisciplinary projects require close coordination among experts who use different technical languages, assumptions, and validation standards. Physicists, numerical researchers, and quantum algorithm developers specialize in different parts of the workflow, making it difficult to communicate intermediate results and maintain consistent conventions across disciplinary boundaries. 

Recent advances in artificial intelligence, particularly multi-agent workflows composed of specialized agents, offer a potential way to integrate expertise across these disciplinary boundaries while enabling automated scientific checks. By combining reasoning with external tools and execution feedback~\cite{Yao2023ReAct}, agents could carry out long-horizon research tasks. Agent-based workflows have recently been introduced to automate complex scientific tasks~\cite{Miao2025PhysMaster,Lu:2025izv,Lu:2026tmf}, including lattice-QCD research~\cite{Tan2026AutoCSK,Gao:2026cpd}, quantum simulations~\cite{Gustin2025QuantumAgent,Zhang:2026qnq,Li:2026jis,Nys:2026oni}, and the design of quantum circuits and codes~\cite{Knipfer:2026kng,Sakka:2026fry,He:2025esj}, such as scalable state-preparation circuits for field theories~\cite{Cao:2025shc}. Guided by independent evaluators or exact verifiers, LLM agents have also found quantum circuits and compilations that improve on previously published constructions~\cite{Gu:2026nkq,Crichigno:2026cbs}, while classically verifiable circuit structure~\cite{Khattar:2026syy} and Lean-based formal verification~\cite{Zhang:2026blk} are being developed to certify such AI-generated results. While these workflows successfully automate tasks within specific domains, 
quantum simulation of QFT requires a workflow that can operate across disciplines.

In this work, we present \texttt{Jianmu-Q},\footnote{In ancient Chinese mythology, Jianmu (\begin{CJK*}{UTF8}{gbsn}建木\end{CJK*}) is a sacred tree that connects heaven and earth, along which deities ascend and descend. The name reflects the role of the workflow in bridging QFT and quantum hardware.} a multi-agent workflow designed to translate a QFT problem into an executable quantum circuit. For a specified model and observable, specialized agents carry out the theoretical formulation, lattice discretization, qubit mapping, and quantum algorithm design (see Fig.~\ref{fig:workflow}). The highlight of this workflow is the combination of derivation review and numerical cross-validation. Independent review agents check whether the derived results are correct, for example, by verifying discretized conservation relations such as $[H_{\text{lat}}, Q_{\text{lat}}] = 0$. Once review agents identify an error, the feedback is sent to the corresponding agent for correction. To verify the correctness of quantum circuits, we execute them on a classical simulator and compare the results with independent calculations from ED and TN. Disagreement with the ED and TN results can automatically trigger further optimization or ansatz redesign, whereas agreement allows the circuit to be submitted to IBM quantum hardware. Our design reduces the risk of errors in derivations or code propagating to the final result, thereby improving the reliability of the workflow. We benchmark \texttt{Jianmu-Q} on light-cone matrix elements in the $(1+1)$-dimensional Nambu--Jona-Lasinio and Schwinger models and on the energy gap of a $(2+1)$-dimensional $\mathbb{Z}_2$ gauge theory: the noiseless simulations converge toward the ED and TN references with increasing circuit depth, and selected circuits are executed on IBM quantum hardware.

This paper is organized as follows:
Section~\ref{sec:physical} introduces the physical tasks in terms of
states, operators, and matrix elements.
Section~\ref{sec:workflow} describes the multi-agent architecture and
its review and cross-validation procedures.
Section~\ref{sec:case} presents the benchmark cases and results.
Section~\ref{discussion} provides a summary and outlook for the workflow.

\section{Physical task}
\label{sec:physical}
Generally, the observables in QFT are given by
\begin{align}
    \mathcal{M} = \sum_{i=1}^n c_i\bra{\phi_i} O_i\ket{\psi_i}\,,
    \label{eq:physical_matrix_element}
\end{align}
where $\{\ket{\psi_i}\}$ and $\{\ket{\phi_i}\}$ are external states and $\{O_i\}$ are arbitrary operators with coefficients $c_i$. Different choices correspond to different physical quantities.

A kind of observable is obtained by choosing $n=1$ and taking both bra and ket states to be the vacuum $\ket{\Omega}$. For example, the quark condensate is obtained by choosing $O_1 = \bar{\psi}(0) \psi(0)$
\begin{align}
    \braket{\bar{\psi}\psi} = \bra{\Omega}\bar{\psi}(0)\psi(0)\ket{\Omega}\,,
\end{align}
where ${\psi}$ is the quark field. This observable is the order parameter of chiral symmetry breaking in QCD. The operator $O_1$ could also be chosen as a time-ordered product, $O_1 = T(\mathcal{O}_1(x_1)\mathcal{O}_2(x_2)\mathcal{O}_3(y_1)...\mathcal{O}_{N+2}(y_N))$, whose vacuum expectation value gives an $(N+2)$-point correlation function. According to the Lehmann–Symanzik–Zimmermann (LSZ) reduction formula, this time-ordered $(N+2)$-point correlation function is related to the scattering amplitude of $h_1(k_1) + h_2(k_2) \to h_3(p_1)+...+h_{N+2}(p_N)$, where $h_\alpha$ is a hadron that has the same quantum numbers as the operator $\mathcal{O}_\alpha$.

Another kind of observable involves non-vacuum states. For example, the dispersion relation of a hadron $h$ is given by
\begin{align}
    E(p) = \bra{h(p)} H \ket{h(p)} - \bra{\Omega} H \ket{\Omega}\,.
\end{align}
Moreover, matrix elements of the energy-momentum tensor encode momentum and angular momentum carried by quarks and gluons~\cite{Ji1997Spin}. The operator can also be chosen as a current-current correlator $J_\mu(x) J_\nu(0)$, whose matrix element is related to the hadronic tensor $\mathcal{W}_{\mu\nu}(q)$,
\begin{align}
  \mathcal{W}_{\mu\nu}(q) = \frac{1}{4\pi} \int d^4x e^{iq\cdot x} \bra{h(p)} J_\mu(x) J_\nu(0) \ket{h(p)}\,.  
\end{align}
The hadronic tensor determines the cross section of deep inelastic scattering.
Through QCD factorization, the hadronic tensor can be factorized into the convolution of the PDFs and partonic hard-scattering coefficients, where the PDFs reveal the internal structure of hadrons. To extract PDFs, the operator $O_i$ in Eq.~\eqref{eq:physical_matrix_element} needs to be non-local. For a quark of flavor $q$, the unpolarized collinear PDF is defined by the light-cone correlation
\begin{equation}
    \begin{split}
    f_{q/h}(x,\mu)
    ={}& \int\frac{dz}{4\pi}\,e^{-ix(p\cdot n)z}\\
       &{}\times\langle h(p)|\bar{\psi}_q(zn)(n\cdot \gamma)      W(zn,0)\psi_q(0)|h(p)\rangle ,
    \end{split}
    \label{eq:physical_pdf}
\end{equation}
where $n = (1,0,0,1)$ and $W(zn,0)$ is the straight Wilson line joining the two quark fields $\bar{\psi}_q(zn)$ and $\psi_q(0)$~\cite{CollinsSoper1982Parton}, $x$ is the longitudinal momentum fraction. The numerical task is to determine this gauge-invariant correlation over a range of separations on quantum computers and then perform the Fourier transform on classical devices.

The bra and ket states of observables can be different. One such case is the light-cone distribution amplitude of a meson state $\ket{h_M(p)}$. The light-cone distribution amplitude is a non-perturbative quantity that enters the description of exclusive processes in QCD. The operator definition of vector-meson LCDAs is given by
\begin{align}\label{eq:defLCDA}
    \phi_h(x,\mu) =& \frac{1}{f_h}\int dz e^{-i(x-1)(n\cdot P)z} \nonumber\\
       &{}\times\langle \Omega|\bar{\psi}_q(zn)(n\cdot \gamma)      W(zn,0)\psi_q(0)|h_M(p)\rangle,
\end{align}
where $f_h$ is the decay constant, defined by $\int_0^1 dx \phi_h = 1$. Therefore, it is necessary for the agent to handle observables with different bra and ket states.

Finite-temperature and finite-density observables introduce another extension. 
At finite temperature and baryon chemical potential, the expectation value takes the form $\operatorname{Tr}(\rho_{\beta,\mu_B}O)$, with $\rho_{\beta,\mu_B}=Z^{-1}\exp[-\beta(H-\mu_B N_B)]$.
Here $N_B$ is baryon number, $\beta$ is inverse temperature, and $Z$ normalizes the density matrix~\cite{Aarts2016FiniteDensity}.
An ensemble decomposition or a purification relates this trace to matrix-element measurements, but preparing the thermal state is an additional task.

These examples show that the observables in Eq.~\eqref{eq:physical_matrix_element} are widely used in high-energy and nuclear physics, including scattering amplitudes, hadron structure, and the QCD phase transition. Thus, the main physical task of \texttt{Jianmu-Q} is to simulate such observables on quantum computers within the framework of LGT. The immediate output is a set of regulated matrix elements on the lattice. Extracting observables in the continuum limit may additionally require renormalization, extrapolation to zero lattice spacing, and extrapolation to infinite volume~\cite{Ji2021LaMET}. However, current noisy intermediate-scale quantum (NISQ) devices are not powerful enough to perform the lattice calculations required for a controlled continuum extrapolation. Renormalization and continuum extrapolation are therefore left to future versions of \texttt{Jianmu-Q}.

\begin{table*}[t]
 \caption{Agents of \texttt{Jianmu-Q} and their tasks. Each agent receives structured input from the preceding stage, and review or validation outcomes are returned to the responsible agent. Agent labels (A1--A10) match those of the public implementation; the noiseless simulation and hardware execution are carried out by deterministic runtimes rather than by LLM agents.}
 \label{tab:agents}
 \begin{ruledtabular}
 \begin{tabular}{p{2.8cm}p{5.2cm}p{8.3cm}}
 Stage & Agent(s) & Task \tabularnewline
 \hline
 \raggedright Theory and mapping & \raggedright physical formulation (A1) with reviewer; lattice formulation (A2) with reviewer; qubit mapping (A3) with reviewer & \raggedright $\mathcal{L}\to H,\,Q$; check $[H,Q]=0$ and external-state quantum numbers; boundary conditions, discretization, digitization; qubit $H$, $O_i$, $Q$ \tabularnewline
 \raggedright Algorithm design & \raggedright state preparation (A4.1); observable measurement (A4.2); Qiskit translation (A5) & \raggedright symmetry-preserving VQE ansatz; measurement and time-evolution circuits; Qiskit adapters \tabularnewline
 \raggedright Simulation & \raggedright noiseless simulation (deterministic runtime) & \raggedright convergence in $D$ and $N_T$; triggers ansatz redesign \tabularnewline
 \raggedright Reference & \raggedright reference calculation: ED (A6), TN (A7) & \raggedright independent solutions of the same lattice problem \tabularnewline
 \raggedright Validation & \raggedright validation (A8) & \raggedright ED--TN agreement; circuit--reference agreement; sources of discrepancies \tabularnewline
 \raggedright Hardware & \raggedright hardware-plan review (A9); hardware execution (deterministic runtime) & \raggedright feasibility assessment and submission decision; transpilation, job submission, error mitigation \tabularnewline
 \raggedright Output & \raggedright results and workflow record (A10) & \raggedright inputs, circuits, results, reasoning and review trace \tabularnewline
 \end{tabular}
 \end{ruledtabular}
\end{table*}

\section{Multi-Agent Workflow}
\label{sec:workflow}
\texttt{Jianmu-Q} is designed to automate the transformation from a QFT problem of the form in Eq.~\eqref{eq:physical_matrix_element} to executable quantum circuits. As depicted in Fig.~\ref{fig:workflow}, the workflow takes the user's concise description as input and concludes with the results of quantum circuit execution. These results are compared with independent ED and TN calculations for numerical cross-validation. To make this transformation reliable, \texttt{Jianmu-Q} uses a modular, multi-agent architecture rather than relying on a single large language model (LLM). In a single-LLM workflow, assumptions made during spatial discretization and qubit mapping can be difficult to track, which makes errors difficult to trace. In contrast, our workflow assigns specific scientific tasks to individual agents that communicate exclusively through structured, machine-readable contracts. In the current implementation, eleven LLM-based agents, labeled A1--A10 with A4 split into A4.1 and A4.2, carry out the scientific tasks, while a deterministic controller and non-LLM runtimes perform the VQE optimization, noiseless simulation, numerical acceptance tests, and hardware execution.
The decisions are separated and explicitly recorded, improving transparency and reducing error propagation. Table~\ref{tab:agents} summarizes the agents and their tasks.
A1 derives the physical formulation of the problem, and A2
constructs the lattice formulation. A3 maps the validated lattice Hamiltonian, observables, and conserved quantities onto qubits. Outputs of A1, A2, and A3 are independently checked by review agents before being passed to the next agent. A4.1 designs state preparation, whereas A4.2 designs observable measurement and time evolution. A5 translates both designs into Qiskit programs; deterministic runtimes optimize the variational parameters and execute the numerical calculations. A6 and A7 generate the independent ED and TN reference programs, A8 reviews the numerical comparisons, A9 assesses hardware execution, and A10 assembles the final report.

The first stage, carried out by the \emph{theory and mapping agents}, is designed to establish the physical problem and its qubit representation. The workflow begins by parsing the continuum problem based on user inputs, which contain the physical model, observables $\{O_i\}$, external states $\{\ket{\psi_i},\ket{\phi_i}\}$, and coefficients $\{c_i\}$. The \emph{physical-formulation agent A1} begins with the Lagrangian $\mathcal{L}$ of the model, then derives the Hamiltonian $H$ and the relevant symmetries, and outputs exact expressions for $H$ and conserved quantities $Q$. An automated symbolic review by a \emph{review agent} then verifies $[H,Q] = 0$ and ensures the quantum numbers of external states align with the model's conservation laws. Once validated, the \emph{lattice-formulation agent A2} discretizes the theory onto a spatial lattice. The agent selects the boundary conditions of the spatial lattice by considering the physical task and the computational resources. For example, the periodic boundary condition (PBC) is preferred when the external state needs to have definite momentum. On the other hand, if translation invariance is not important to the input problem, agent would choose open boundary conditions (OBCs), which could reduce the computational cost for both TN and quantum computing calculations. Due to the Nielsen–Ninomiya theorem~\cite{Nielsen:1981hk}, lattice fermion discretization generally requires trade-offs among continuum symmetries and other desired properties. The agent selects a discretization scheme for fermions that preserves the symmetry relevant to the physical task. After discretization, the gauge-field Hilbert space is truncated through a digitization scheme. The agent would select an appropriate digitization method from existing schemes~\cite{Bauer:2022hpo}. The procedure for choosing boundary conditions, discretization, and digitization of quantum fields is guided by a dynamically retrieved literature~\cite{Smit2002Lattice, Gattringer2010Lattice}, followed by an embedded algebraic review to ensure consistency. Finally, the \emph{qubit-mapping agent A3} maps lattice quantum fields to qubits, such as applying the Jordan-Wigner transformation~\cite{Jordan1928Pauli} for fermions and binary encoding for gauge fields~\cite{Shaw:2020udc}. The Hamiltonian, observables, and conserved charges are then obtained from the corresponding qubit representations of the elementary field operators.

The next stage, carried out by the \emph{quantum-algorithm-design agents}, is to design a quantum algorithm circuit to simulate the observables. State preparation and observable measurement are designed in parallel by a \emph{state-preparation agent A4.1} and an \emph{observable-measurement agent A4.2}. In most cases, the external states $\{\ket{\psi_i},\ket{\phi_i}\}$ are vacuum and hadron states, which are eigenstates with specified conserved quantum numbers $Q$. Thus, we let A4.1 use a variational quantum eigensolver (VQE) to prepare the external state. VQE can use relatively shallow parameterized circuits, making it suitable for NISQ devices. The choice of initial reference states and the design of the ansatz are two key procedures of VQE. We let the agent choose the initial reference state that has the same quantum number as the target external state, and the ansatz $U(\theta)$ is designed such that it satisfies $[U(\theta),Q] = 0$. These requirements keep the trial state in the same symmetry sector as the target state. The VQE loss function is constructed for the target state and minimized by a classical optimizer. A4.2 then designs quantum circuits, such as Hadamard tests and circuits for dynamical correlation functions, to measure $\bra{\phi_i}O_i\ket{\psi_i}$ for all $i$~\cite{Pedernales:2014izf,z126-zdqj}. The quantum circuit for simulating time evolution is also designed if the observable $O_i$ depends on time. This can be done by the Trotter algorithm or quantum singular value transformation (QSVT). All circuits are implemented in Qiskit and transpiled for execution on IBM quantum hardware~\cite{JavadiAbhari2024Qiskit}.

The \emph{Qiskit translation agent A5} translates the designed circuits into Qiskit programs. Owing to the limitations of current quantum hardware, these programs are first executed by a deterministic \emph{noiseless-simulation runtime} on a classical simulator before any submission to IBM quantum devices.
At each circuit depth, the VQE parameters are optimized by minimizing the loss function~\cite{Peruzzo2014VQE}. In the noiseless simulation, 
we test the ansatz at depth $D = 1,2,4,...$ and monitor how the optimized energy and energy variance change with increasing depth. The optimized parameters at one depth are used to initialize the next computation. If the energy error and variance decrease with increasing ansatz depth $D$ and eventually fall below the thresholds, the ansatz is considered efficient for the problem.
If the state preparation does not improve as $D$ increases, the ansatz is considered inefficient for the problem and is redesigned by the \emph{quantum-algorithm-design agents} (the \emph{ansatz-redesign} loop). The number of redesign attempts is bounded. Inefficient ansatz will not be selected when the agent encounters similar problems next time. All attempts, including unsuccessful ones, are retained to keep the selection process traceable. Once the states are prepared, the simulator evaluates the observable in Eq.~\eqref{eq:physical_matrix_element} using the designed circuit. If the operators $O_i$ depend on time and the Trotter algorithm is used to simulate time evolution, the deterministic \emph{noiseless-simulation runtime} tests how the results converge as the number of Trotter steps $N_T$ increases.

In parallel with the circuit simulation, the \emph{reference-calculation agents} construct two independent solutions of the same lattice problem. 
For the same small system used in the earlier circuit tests, the ED program generated by A6 uses dense or sparse diagonalization via SciPy~\cite{Virtanen2020SciPy}. 
The TN program generated by A7 uses matrix-product states and operators with ITensor~\cite{White1992DMRG,Fishman2022ITensor}.
The classical algorithms receive the validated lattice Hamiltonian, states, and observables, but do not read the circuit and its results on a classical simulator. The deterministic cross-check runtime compares the ED and TN numerical results. It then compares the noiseless simulation result with them. The comparison includes both absolute differences and convergence with increasing $D$ and $N_T$. For any discrepancies between these results, A8 identifies possible sources, including state-preparation error, Trotter error, and TN truncation.

These numerical cross-checks are essential because successful circuit execution alone does not establish the accuracy of the target observable. Appendix~\ref{app:failed_crosscheck} illustrates this point with a Schwinger-model calculation: ED and TN give closely matching matrix elements, whereas the noiseless circuit result differs substantially at nonzero separations. Independent references thus reveal a discrepancy that circuit execution alone would not identify.

Execution on a quantum device is optional and is implemented by a deterministic hardware runtime, with scientific feasibility assessed by the \emph{hardware-plan review agent A9}. The runtime transpiles candidate circuits into the native instruction set of each candidate backend and records circuit width, depth, two-qubit gate count, calibration data, and available execution time. A9 reviews this evidence to assess whether a meaningful physical signal is likely to survive hardware noise and recommends either submission of selected circuit--backend pairs or no submission. The runtime enforces this decision and handles job submission, result retrieval, and the configured error-mitigation procedures, including measurement-error mitigation and zero-noise extrapolation~\cite{Temme2017ZNE,Li2017Mitigation}. The hardware result is compared with the noiseless simulation result of the same circuit. Given the current limitations of NISQ devices, this comparison is reported as a device test and does not replace the convergence study in $D$ and $N_T$ performed using classical simulation.

The final output, the \emph{results and workflow record}, is assembled by the report agent A10. The record collects the input model, the validated observables, the designed circuit, classical results, and quantum device results (when requested). It also retains the reasoning and review record that produced them, so that the entire sequence of model-generated decisions is traceable.

\section{Benchmarks and Results}
\label{sec:case}

To test \texttt{Jianmu-Q} across different physical settings, we present three representative benchmarks. These benchmarks increase in complexity from a purely fermionic model to LGTs in one and two spatial dimensions. The agent labels below follow Table~\ref{tab:agents}. The benchmarks reported below use GPT-5.6 Sol. In separate supplementary runs using GPT-6 Sol, the task specifications contain approximately 181--374 user-input tokens, with at most 16 additional clarification tokens, measured using the fixed \texttt{o200k\_base} tokenizer. These counts describe the length of the supplied scientific inputs and are independent of the execution model for unchanged inputs. The supplementary runs involve 16--21 LLM calls; their cumulative input and output token usage, which depends on the model and execution history, is reported in Appendix~\ref{app:resource_usage}.

\subsection{1+1D NJL model}
\label{sec:benchmark_njl}
We first consider a purely fermionic example: the one-flavor Nambu--Jona-Lasinio (NJL) model in $1+1$ dimensions~\cite{Nambu1961NJL,Gross1974GN,Hatsuda:1994pi},
\begin{equation}
 \mathcal L_{\rm NJL}=\bar\psi(i\gamma^\mu\partial_\mu-m)\psi
 +g(\bar\psi\psi)^2 .
 \label{eq:njl_lagrangian}
\end{equation}
From $\mathcal L_{\rm NJL}$, A1 derives the continuum Hamiltonian and identifies the conserved global $U(1)$ charge, which defines the neutral sector used below.
NJL model at finite density has also been studied with tensor networks~\cite{Akiyama:2020soe}. The input parameters are $m=0.5$, $g = 0.5$, lattice spacing $a=1$, and $L=5$ spatial sites labelled by $z=-2,-1,0,1,2$. The target is the vacuum-subtracted light-cone correlator, which is the position-space correlator entering the PDF definition in Eq.~\eqref{eq:physical_pdf}
\begin{equation}
 \begin{aligned}
 \mathcal M_{\rm NJL}(z) &= \langle h|O(z)|h\rangle-\langle\Omega|O(z)|\Omega\rangle,\\
 O(z)&=\bar\psi(zn)(n\cdot \gamma)\psi(0),
 \end{aligned}
 \label{eq:njl_observable}
\end{equation}
where $n^\mu=(1,1)$ in $1+1$ dimensions, so $n\cdot \gamma = \gamma^-$. Here $|\Omega\rangle$ and $|h\rangle$ are the lowest and first excited states in the neutral global-$U(1)$ charge sector, respectively.  Since the NJL model in Eq.~\eqref{eq:njl_lagrangian} contains no gauge field, the Wilson line reduces to the identity, and no gauge-field truncation, or Gauss-law constraint is required.

\begin{figure}[tp]
 \centering
 \includegraphics[width=1\columnwidth]{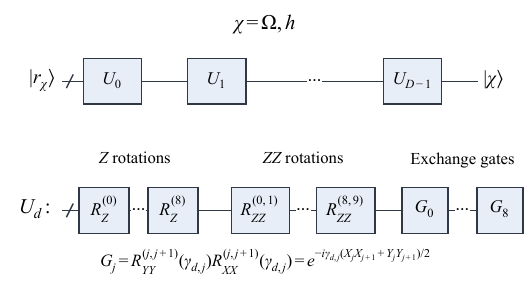}
 \caption{Ten-qubit state-preparation ansatz $V_D(\theta) = U_{D-1}...U_1U_0$ for the NJL model. The upper circuit represents the same $D$-layer ansatz applied to the two orthogonal reference states. After optimization, the two reference states are mapped to the vacuum $|\Omega\rangle$ and hadron state $|h\rangle$.
 The vacuum reference state is given by $\ket{r_\Omega} = [(\ket{01}-\ket{10})/\sqrt{2}]^{\otimes 5}$, while $\ket{r_h}$ is obtained by replacing the central $(|01\rangle-|10\rangle)/\sqrt{2}$ with $(|01\rangle+|10\rangle)/\sqrt{2}$. 
 The lower circuit shows one ansatz layer $U_d, (d=0,1…,D-1)$, consisting of $R_Z$, $R_{ZZ}$, and exchange-gate layers. The superscripts label the qubit indices, and ellipses denote the omitted gates or layers in ascending qubit order.}
 \label{fig:njl_ansatz}
\end{figure}

A2 selects Wilson fermions to discretize the fermion field on the lattice, with Wilson coefficient $r=1$ and OBCs. Although OBCs violate the translation symmetry, the $\ket{h}$ state can still have a good approximation to the zero momentum state at large enough lattice volume. With $\gamma^0=\sigma^1$, $\gamma^1=-i\sigma^2$, and $\alpha^1=\gamma^0\gamma^1=\sigma^3$, the resulting finite-lattice Hamiltonian is
\begin{equation}
 \begin{aligned}
 H_{\rm NJL}={}&\sum_{z=-2}^{2}\left[
 \left(m+\frac{r}{a}\right)c_z^\dagger\sigma^1 c_z
 -\frac{g}{a}(c_z^\dagger\sigma^1 c_z)^2\right]\\
 &+\sum_{z=-2}^{1}\left[
 c_z^\dagger\left(-\frac{i\sigma^3}{2a}
                  -\frac{r\sigma^1}{2a}\right)c_{z+1}
 +\mathrm{H.c.}\right] .
 \end{aligned}
 \label{eq:njl_hamiltonian}
\end{equation}
The conserved lattice charge is
\begin{align}
    &Q = \sum_{z=-2}^2 \mathcal{Q}_z=\sum_{z=-2}^{2}\sum_{s=0}^{1}c_{z,s}^\dagger c_{z,s}-L,\nonumber\\
    &[H_{\rm NJL},Q]=0 .
 \label{eq:Qz}
\end{align}

A3 maps the ten fermion modes to ten qubits by the Jordan--Wigner transformation, ordered as $q(z,s)=2(z+2)+s$,
\begin{equation}
 c_q=\left(\prod_{q'<q}\sigma^3_{q'}\right)\frac{\sigma^1_q+i\sigma^2_q}{2} \equiv \Xi_q \frac{\sigma^1_q+i\sigma^2_q}{2}.
\end{equation}
Using $\gamma^0\gamma^-=1 -\sigma^3$, the regulated bilocal operator is
\begin{equation}
 \begin{aligned}
 O(z) &= e^{iH_{\rm NJL}z} c_z^\dagger e^{-iH_{\rm NJL}z} (1 - \sigma^3)c_0 \\
 &= \sum_{\alpha,\beta=1}^2 f_{\alpha\beta} e^{iH_{\rm NJL}z} \mathcal{O}_\alpha(z) e^{-iH_{\rm NJL}z} \mathcal{O}_\beta\,,
 \end{aligned}
 \label{eq:njl_lattice_observable}
\end{equation}
where
\begin{align}
    &\mathcal{O}_\alpha(z) = \Xi_{2z+5} \sigma^\alpha_{2z+5}\,,\nonumber\\
    &\mathcal{O}_\beta = \Xi_5 \sigma^\beta_5\,.
\end{align}
$f_{\alpha\beta}$ is given by $f_{11} = f_{22} = 1/2$, and $f_{12} = -f_{21} = i/2$.

To prepare the two external states simultaneously, A4.1 selects a symmetry-preserving subspace-search VQE (SSVQE)~\cite{Nakanishi2019SSVQE}. Two orthogonal zero-charge reference states $\ket{r_\Omega}$ and $\ket{r_h}$ are acted on by the same ansatz $V_D(\theta)$, where $D$ is the ansatz depth and $\theta$ denotes all parameters in the circuit. The quantum circuit for the ansatz is shown in Fig.~\ref{fig:njl_ansatz}. Each layer contains nearest-neighbor $R_Z$ and $R_{ZZ}$ rotations followed by shared-angle $R_{XX}R_{YY}$ two-qubit rotations, giving 27 parameters per layer. Every layer commutes with $Q$ and therefore preserves the zero-charge sector and fermion parity. The loss function of VQE is given by
\begin{align}
     \mathcal{C}(\theta)=&\frac23\langle r_\Omega|V_D^\dagger(\theta) H_{\rm qc}V_D(\theta)|r_\Omega\rangle \nonumber\\
     &+\frac13\langle r_h|V_D^\dagger(\theta) H_{\rm qc}V_D(\theta)|r_h\rangle\,.
 \label{eq:njl_ssvqe_loss}
\end{align}
A5 implements the ansatz in Qiskit, and the VQE runtime minimizes this loss function at $D=1,2,4,6$ with warm starts, using the optimized parameters at one depth to initialize the next. The optimized parameters are denoted as $\theta^*$. The optimized states are therefore $\ket{\Omega} = V_D(\theta^*) \ket{r_\Omega}$ and $\ket{h} = V_D(\theta^*)\ket{r_h}$. The optimized vacuum and hadron energies at different $D$ are listed in Tab.~\ref{tab:njl_energies}.

The independent reference programs generated by A6 and A7 yield the ED and TN energies $E_\Omega$ and $E_h$, shown in the first and second rows of Tab.~\ref{tab:njl_energies}.
The agreement between ED and TN provides the independent reference for validating the quantum-circuit calculation.
The cross-check runtime compares the noiseless simulation results with these independent references, and A8 reviews the resulting discrepancies and convergence diagnostics.
At $D=1$, the vacuum and hadron energy errors are $7.62\times10^{-2}$ and $1.11$, with variances $2.94\times10^{-1}$ and $8.76\times10^{-1}$, respectively. At $D=6$, the corresponding energy errors decrease to $6.95\times10^{-4}$ and $1.17\times10^{-3}$, while the variances decrease to $3.33\times10^{-3}$ and $4.35\times10^{-3}$. 
The systematic decrease in both energy error and variance shows that the ansatz is efficient for this problem. The required depth depends on the target precision.

\begin{figure}[t]
 \centering
 \includegraphics[width=1\columnwidth]{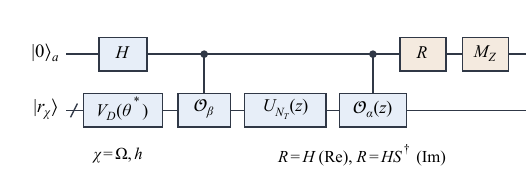}
 \caption{Quantum circuit for measuring the NJL light-cone correlator. The external state $\ket{\chi}$, with $\chi=\Omega, h$, is prepared from the reference state $\ket{r_{\chi}}$ using the optimized ansatz $V_D(\theta^*)$. The controlled operators $\mathcal{O}_{\beta}$ and $\mathcal{O}_{\alpha}(z)$, together with the Trotterized time evolution $U_{N_T}(z)$, measure each Pauli-string contribution in Eq.~\eqref{eq:njl_lattice_observable}. The real and imaginary parts are obtained by choosing $R=H$ and $R=HS^\dagger$, respectively, before measuring the ancilla in the computational basis.}
 \label{fig:njl_circuit}
\end{figure}

In parallel with state preparation, A4.2 designs the measurement of $\mathcal{M}_{\rm NJL}(z)$. For $z\ne0$, $U(z)=e^{-iH_{\rm NJL}z}$ is approximated using a second-order Suzuki product formula with $N_T=1,2,4,8,16$ Trotter steps~\cite{Suzuki1990ProductFormula}, which is denoted by $U_{N_T}(z)$. As shown in Eq.~\eqref{eq:njl_lattice_observable}, $O(z)$ is decomposed into four Pauli-string contributions for each external state. A5 implements this measurement design as the Qiskit interferometer in Fig.~\ref{fig:njl_circuit}, which uses one ancilla, controlled Pauli strings corresponding to the fermion fields, and a common uncontrolled time evolution. The real and imaginary parts of $\mathcal{M}_{\rm NJL}(z)$ can be obtained by applying $R = H$ and $R = HS^\dagger$, respectively, before measuring the ancilla qubit. At $z=0$, the direct operator $O_{\rm qc}(0)=I-\sigma^3_5$ is measured without time evolution.

\begin{table}[t]
 \caption{External-state energies in lattice units from ED, TN, and noiseless simulations (NS) at different ansatz depths $D$. The IBM results use $D=1$ and include measurement mitigation and zero-noise extrapolation (ZNE). The uncertainties are propagated statistical errors.}
 \label{tab:njl_energies}
 \begin{ruledtabular}
 \begin{tabular}{lrrr}
 Method & $E_\Omega$ & $E_h$ & $\Delta E$\\
 \hline
ED & $-10.508406$ & $-8.506187$ & $2.002219$\\
TN & $-10.508406$ & $-8.506187$ & $2.002219$\\
NS, $D=1$ & $-10.432190$ & $-7.392336$ & $3.039854$\\
NS, $D=2$ & $-10.482874$ & $-8.290420$ & $2.192453$\\
NS, $D=4$ & $-10.506781$ & $-8.497024$ & $2.009757$\\
NS, $D=6$ & $-10.507711$ & $-8.505016$ & $2.002694$\\
IBM, $D=1$& $-9.985 \pm 0.031$ & $-7.103 \pm 0.031$ & $2.883 \pm 0.044$\\
 \end{tabular}
 \end{ruledtabular}
\end{table}

\begin{figure}[t]
 \centering
 \includegraphics[width=1\columnwidth]{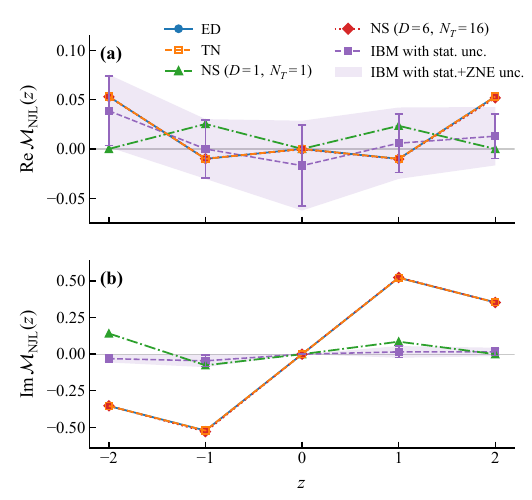}
 \caption{NJL light-cone correlator $\mathcal{M}_{\rm NJL}(z)$. (a) and (b) show the real and imaginary parts, respectively. Blue circles and orange squares denote ED and TN, providing the independent references. Green triangles and red diamonds are noiseless simulation (NS) results at $(D,N_T)=(1,1)$ and $(6,16)$, respectively. Purple squares are the IBM results obtained from $(D,N_T)=(1,1)$ circuits and should therefore be compared with the corresponding green NS curve. The purple error bars show statistical uncertainties, while the light-purple band additionally includes the uncertainty from the ZNE procedure.}
 \label{fig:njl_benchmark}
\end{figure}

The green and red curves in Fig.~\ref{fig:njl_benchmark} show that the noiseless simulation approaches the independent references as $D$ and $N_T$ are increased, from $(1,1)$ up to $(6,16)$. At $(D,N_T)=(6,16)$, the absolute complex errors relative to ED at $z=-2,...,2$ are at most $7.33\times10^{-3}$, indicating good agreement with the ED result. Further improvement of the noiseless simulation result can be obtained by increasing $D$ and extrapolating to zero Trotter error using the results at $N_T = 1,2,4,8,16$.

For the hardware demonstration, A9 assesses the compiled circuit resources and backend information before the hardware runtime submits the $(D,N_T)=(1,1)$ circuits to \texttt{ibm\_kawasaki}. The retrieved results give the five purple data points with error bars.
The corresponding noiseless reference is therefore the green curve.
The circuit depths of $\mathcal{M}_{\rm NJL}(z\not=0)$ are $2033$--$2077$, with approximately $1.44\times10^4$ two-qubit gates. The $z=0$ circuits are much smaller, with depth 101 and 78 two-qubit gates in total.

Measurement mitigation and component-wise zero-noise extrapolation (ZNE) with noise factors $1,3,5$ are attempted~\cite{Temme2017ZNE}. Stable components use linear or exponential extrapolation, whereas unstable fits revert to the mitigated noise-factor-one value. The purple error bars in Fig.~\ref{fig:njl_benchmark} propagate the selected components' statistical errors. The light-purple band additionally propagates the spread among admissible linear, exponential, and fallback estimates as a ZNE-model uncertainty. 
A deep hardware circuit can yield a statistically precise but systematically suppressed signal, explaining why several purple points remain inconsistent with the corresponding noiseless results even after mitigation. Although $(D,N_T)=(1,1)$ is the shallowest setting considered, its $z\neq0$ circuits still have depths above 2000. This submission is therefore an exploratory hardware test, and quantitative agreement with the corresponding NS curve is not expected at current noise levels.

This benchmark demonstrates the end-to-end workflow from a continuum field theory to sector-preserving state preparation, real-time measurement circuits, independent classical checks, and hardware execution.
It also records the limits of the calculation: the hardware data constitute a shallow-circuit demonstration rather than a quantitative prediction.

\subsection{Schwinger model}
\label{sec:Schwinger}
The second benchmark considers the Schwinger model, which couples a fermion field to a $U(1)$ gauge field in one spatial dimension~\cite{Schwinger1962}. The task starts from the Lagrangian
\begin{equation}
 \mathcal L=-\frac14 F_{\mu\nu}F^{\mu\nu}
 +\bar\psi(i\gamma^\mu D_\mu-m)\psi,\
 D_\mu=\partial_\mu+igA_\mu .
\end{equation}
From this Lagrangian, A1 derives the continuum Hamiltonian and identifies the conserved electric charge and Gauss-law constraint.
The input fixes bare quark mass $m = 0.5$, bare coupling $g = 0.5$, lattice spacing $a=1$, and the number of spatial sites $L=5$. The requested observable is the light-cone matrix element entering the LCDA definition in Eq.~\eqref{eq:defLCDA}
\begin{equation}
 \mathcal M_{\rm LCDA}(z)=\langle\Omega|\bar\psi(zn)\gamma^-
 W(zn,0)\psi(0)|h\rangle,
 \label{eq:schwinger_target}
\end{equation}
where $n^\mu=(1,1)$, $\gamma^-=\gamma^0-\gamma^1$, and $z=-2,-1,0,1,2$. The Wilson line $W(zn,0)=\mathcal P\exp[-ig\int_0^z ds (n\cdot A(sn))]$ connects the two fermion fields and ensures gauge invariance. The external states $|\Omega\rangle$ and $|h\rangle$ are the lowest and first excited states in the zero-charge sector. Computing Eq.~\eqref{eq:schwinger_target} therefore requires both state preparation and real-time evolution of a gauge-invariant operator.

To discretize the theory, A2 selects OBCs and uses two-component Wilson fermions with Wilson parameter $r=1$. The labels of lattice sites are $z = -2,-1,0,1,2$. To treat the gauge field, A2 chooses the spatial axial gauge, $A_1=0$, allowing the electric fields $E_\ell$ to be eliminated through Gauss's law,
\begin{align}
    E_\ell&=g\left(\sum_{z=-2}^\ell \mathcal{Q}_z -\frac Q2\right).
\end{align}
The definitions of $Q$ and $\mathcal{Q}_z$ are the same as in Eq.~\eqref{eq:Qz}. Here $\ell=-2,\ldots,1$ labels the interior links, and the boundary fluxes are $E_{-3}=-gQ/2$ and $E_2=gQ/2$. The Hamiltonian includes only the electric-field energy on the bulk links, excluding the boundary contribution. The energy of the electric field is therefore expressed as a Coulomb interaction between the quarks. With $\gamma^0=\sigma_1$ and $\gamma^1=-i\sigma_2$, the resulting Hamiltonian is
\begin{align}
 H_{\rm Sch}={}&\sum_{z=-2}^{2}\left(m+\frac ra\right)c_z^\dagger\sigma^1c_z
 \nonumber\\
 &+\sum_{z=-2}^{1}\left[c_z^\dagger
 \left(-\frac{i\sigma^3}{2a}-\frac{r\sigma^1}{2a}\right)c_{z+1}
 +\mathrm{H.c.}\right]\nonumber\\
 &+\frac{g^2}{2}\sum_{\ell=-2}^{1}\left(\sum_{z=-2}^\ell \mathcal{Q}_z -\frac Q2\right)^2 .
 \label{eq:schwinger_wilson_hamiltonian}
\end{align}
A3 then maps the fermion fields to qubits through the Jordan--Wigner transformation.

The next step is to determine the Wilson line. Setting $A_1=0$ leaves the temporal gauge potential along the light-cone path. A2 reconstructs this potential as
\begin{equation}
 A_{0}(z)=g\sum_{\ell=-2}^{1}
 \left(\mathbf \theta_{z,\ell}-\frac12\right)
 \left(\sum_{x=-2}^\ell \mathcal{Q}_x-\frac Q2\right)\,,
\end{equation}
where $\theta_{z,\ell} = 1$ when $z\leq \ell$ and $\theta_{z,\ell} = 0$ when $z>\ell$. In general, the coefficient should be $\theta_{z,\ell}+c$. Here we set $c=-1/2$ to preserve the parity symmetry. The gauge link at spacetime point $(y,y)$ is
\begin{align}
    \mathcal{U}_y = e^{iH_{\mathrm{Sch}}y}\exp(-igA_0(y)) e^{-iH_{\mathrm{Sch}}y}\,.
\end{align}
Thus Wilson line $W(zn,0)$ is given by
\begin{align}
    W(zn,0) = 
    \begin{cases}
        \mathcal{U}_{z-1}...\mathcal{U}_0\,,\ z>0\,,\\
        I,\ z=0\ ,\\
        \mathcal{U}^\dagger_z...\mathcal{U}^\dagger_{-1}\,,\ z<0\,.
    \end{cases}
\end{align}
The observable becomes
\begin{equation}
 \mathcal M_{\rm LCDA}(z) = 2 \langle\Omega|
 e^{iH_{\mathrm{Sch}}z} c_{z,1}^\dagger e^{-iH_{\mathrm{Sch}}z} W(zn,0) c_{0,1}|h\rangle .
 \label{eq:schwinger_lattice_observable}
\end{equation}
The second spinor component is selected by $\gamma^0\gamma^-=1-\sigma^3$.

\begin{figure}[tp]
 \centering
 \includegraphics[width=1\columnwidth]{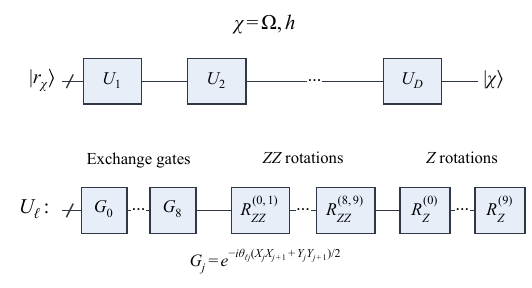}
 \caption{State preparation ansatz for the Schwinger model. Each horizontal line represents the ten-qubit register. The upper circuit applies $D$ layers of the ansatz $V_D(\theta) = U_D ... U_2 U_1 $ to each reference state $|r_\chi\rangle$, with $\chi=\Omega,h$. Superscripts on the rotation gates specify the qubit indices on which they act, and each gate has an independent angle in each layer. The exchange gates $G_j$ and the $R_{ZZ}$ gates act on adjacent qubits in ascending order of $j$, followed by single-qubit $R_Z$ rotations. Each layer has 28 independent parameters and preserves the total electric charge $Q$.}
 \label{fig:schwinger_ansatz}
\end{figure}

To prepare the two external states in the same $Q = 0$ sector, A4.1 chooses a shared variational circuit $V_D(\theta)$ acting on two orthogonal reference states in the $Q = 0$ subspace, which are denoted by $\ket{r_\chi}$ with $\chi = \Omega,h$. Each layer contains nine nearest-neighbor exchange gates, nine $R_{ZZ}$ rotations, and ten $R_Z$ rotations, as shown in Fig.~\ref{fig:schwinger_ansatz}. 
The ansatz preserves $Q$, while the shared unitary preserves the orthogonality of the two reference states. A5 implements this ansatz in Qiskit, and the VQE runtime optimizes the 28 parameters per layer with the weighted subspace-search VQE loss function~\cite{Nakanishi2019SSVQE}
\begin{equation}
 \mathcal C(\theta)=\frac{2E_\Omega(\theta)+E_h(\theta)}{3}\,.
\end{equation}
Here $E_\Omega(\theta)$ and $E_h(\theta)$ are the energy expectation values of the two variational states.
The larger weight selects $\ket{\Omega}$ as the lower energy state. The calculation scans ansatz depths $D=1,2,4,8$, using the optimized parameters at one depth to initialize the next.

\begin{figure}[tp]
 \centering
 \includegraphics[width=1\columnwidth]{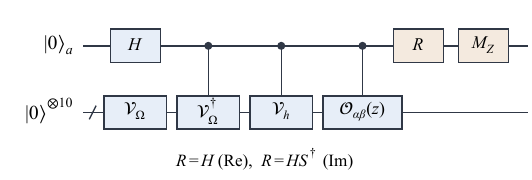}
 \caption{Measurement circuit for $\mathcal{M}_{\alpha\beta}(z)$ in Eq.~\eqref{eq:schwinger_unitary_terms}. The lower line represents the ten-qubit register. The state-preparation circuits satisfy $\mathcal{V}_h\ket{0}^{\otimes10}= V_D(\theta^*) \ket{r_h} = \ket{h}$ and $\mathcal{V}_\Omega \ket{0}^{\otimes10}= V_D(\theta^*)\ket{r_\Omega} = \ket{\Omega}$. The controlled operations prepare $(\ket{0}\ket{\Omega}+\ket{1} \mathcal{O}_{\alpha\beta}\ket{h})/\sqrt2$. Measuring $Z_a$ after $R=H$ gives $\operatorname{Re}\mathcal{M}_{\alpha\beta}$, while $R=HS^\dagger$, corresponding to $S^\dagger$ followed by $H$, gives $\operatorname{Im}\mathcal{M}_{\alpha\beta}$. The block $\mathcal{O}_{\alpha\beta}(z)$ contains the Pauli strings, time evolution, and Wilson line defined in Eq.~\eqref{eq:schwinger_unitary_terms}, with the the time-evolution operators approximated by the second-order product formula.}
 \label{fig:schwinger_measurement}
\end{figure}

The measurement stage must evaluate a transition amplitude between these two states. Expanding the fermion operators gives four unitary matrix elements,
\begin{align}
 \mathcal{M}_{\alpha\beta}(z)&=\langle\Omega|\mathcal{O}_{\alpha\beta}(z)|h\rangle,\nonumber\\
 \mathcal{O}_{\alpha\beta}(z)&= e^{i H_{\rm Sch} z} \Xi_{2(z+2)+1}\sigma^\alpha_{2(z+2)+1} e^{-iH_{\rm Sch}z} \nonumber\\
 &\quad\times W(zn,0) \Xi_{5}\sigma^\beta_5,
 \label{eq:schwinger_unitary_terms}
\end{align}
where $\alpha,\beta = 1,2$. The required combination is
\begin{equation}
 \mathcal M_{\rm LCDA}(z)=\frac{\mathcal{M}_{11}+i\mathcal{M}_{12}-i\mathcal{M}_{21}+\mathcal{M}_{22}}{2}.
\end{equation}
A4.2 designs the Hadamard-test circuit in Fig.~\ref{fig:schwinger_measurement} to measure the real and imaginary parts of each term, and A5 translates it into Qiskit. This requires eight circuits for a given $z$ in $\mathcal{M}_{\rm LCDA}(z)$, each with ten system qubits and one ancilla qubit. Time evolution is implemented using a symmetric second-order product formula that preserves the total charge $Q$.

\begin{figure}[tp]
 \centering
 \includegraphics[width=1\columnwidth]{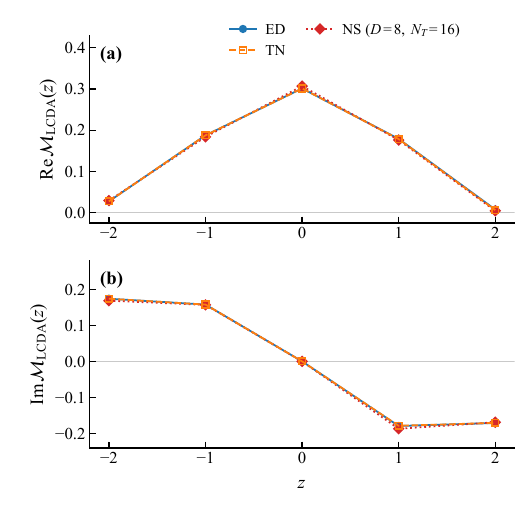}
 \caption{Real (a) and imaginary (b) parts of the light-cone transition matrix element $\mathcal{M}_{\rm LCDA}(z)$. The ED and TN results are compared with the NS result at $(D,N_T)=(8,16)$. The overall phase of each method is fixed by requiring $\operatorname{Im}\mathcal{M}_{\rm LCDA}(0)=0$ and $\operatorname{Re}\mathcal{M}_{\rm LCDA}(0)>0$. For each method, the same phase rotation is applied to all five values of $z$. }
 \label{fig:schwinger_observable}
\end{figure}

The independent reference programs generated by A6 and A7 evaluate the same Hamiltonian and observable using ED with SciPy~\cite{Virtanen2020SciPy} and TN methods with ITensor~\cite{Fishman2022ITensor}, respectively. Tab.~\ref{tab:schwinger_energies} compares their energies with the noiseless simulation results. In the numerical cross-check, the ED and TN energies agree within $5\times10^{-12}$, while the $D=8$ variational energy errors are $2.69\times10^{-4}$ and $1.51\times10^{-3}$ for $|\Omega\rangle$ and $|h\rangle$. The shared ansatz maintains their orthogonality, with squared overlap $1.6\times10^{-32}$.

\begin{table}[tbp]
 \caption{External-state energies for the five-site Schwinger model at $g=m=0.5$ and $a=1$, obtained from ED, TN, and NS at different ansatz depths $D$.}
 \label{tab:schwinger_energies}
 \begin{ruledtabular}
 \begin{tabular}{lrrr}
 Method & $E_\Omega$ & $E_h$ & $\Delta E$\\
 \hline
 ED & $-8.146728$ & $-6.581876$ & $1.564852$\\
 TN & $-8.146728$ & $-6.581876$ & $1.564852$\\
 NS, $D=1$ & $-7.779048$ & $-4.971685$ & $2.807362$\\
 NS, $D=2$ & $-8.057983$ & $-5.750662$ & $2.307321$\\
 NS, $D=4$ & $-8.140596$ & $-6.541190$ & $1.599406$\\
 NS, $D=8$ & $-8.146459$ & $-6.580367$ & $1.566092$\\
 \end{tabular}
 \end{ruledtabular}
\end{table}

Because the bra and ket states of $\mathcal{M}_{\rm LCDA}(z)$ are different, eigenstates can carry different unphysical relative phases, leading to an overall unphysical phase difference in $\mathcal{M}_{\rm LCDA}(z)$. We therefore align this phase before comparing different methods. Parity and charge-conjugation symmetries imply ${\rm Im}\mathcal{M}_{\rm LCDA}(0) = 0$. We rotate each result by a constant phase such that ${\rm Im}\mathcal{M}_{\rm LCDA}(0) = 0$ and ${\rm Re} \mathcal{M}_{\rm LCDA}(0) >0$.
The results of $\mathcal{M}_{\rm LCDA}(z)$ are shown in Fig.~\ref{fig:schwinger_observable}. The maximum difference from ED decreases from $0.594$ at $(D,N_T)=(1,1)$ to $8.06\times10^{-3}$ at $(8,16)$.

Finally, the hardware-assessment stage (A9 and the hardware runtime) evaluates whether the $D=1$, $N_T=1$ measurement circuits can be executed on \texttt{ibm\_kawasaki}. At nonzero separations, the transpiled circuit depths range from 7856 to 17389, exceeding the configured limit of 5000. The $z=0$ circuits have a smaller maximum depth of 3352 but still exceed the calibration-based two-qubit error budget. Hardware submission is therefore skipped, and this benchmark ends with the noiseless simulation and independent reference results.

\subsection{$2+1$D $\mathbb{Z}_2$ gauge theory}
\label{sec:2+1DZ2}
The third benchmark considers a pure $\mathbb{Z}_2$ gauge theory in two spatial dimensions. $\mathbb{Z}_2$ LGTs are common testbeds for quantum simulation~\cite{Yamamoto:2020eqi,Hayata:2024smx,Azad:2025afj,Sekiyama:2026rgt,Nagano:2026fuh} and have also been studied with tensor networks~\cite{Kuramashi:2018mmi}. The benchmark starts from the lattice Hamiltonian~\cite{Wegner1971Duality,KogutSusskind1975}
\begin{equation}
    H=-\frac{1}{a}\left(g^2\sum_\ell X_\ell + g^{-2}\sum_p B_p\right),\ B_p=\prod_{\ell\in\partial p}Z_\ell ,
    \label{eq:z2_link_hamiltonian}
\end{equation}
where $\ell$ and $p$ label links and plaquettes, respectively, and $\partial p$ denotes the boundary of plaquette $p$. The input fixes the bare coupling $g=1$, lattice spacing $a = 1$, and a $4\times4$ lattice. Starting from this Hamiltonian, A1 identifies the physical zero-charge sector through Gauss's law at each vertex $v$,
\begin{align}\label{eq:GauZ2}
    G_v \ket{\rm Phys} = \prod_{\ell\ni v} X_\ell \ket{\rm Phys}=  \ket{\rm Phys} 
\end{align}
The requested observable is 
\begin{align}\label{eq:z2o}
    \Delta E=E_G-E_\Omega\,,
\end{align}
where $E_\Omega$ is the energy of the vacuum state $\ket{\Omega}$, and $E_G$ is the energy of the ``glueball'' state $\ket{G}$ in $Z_2$ gauge theory. $\ket{G}$ is defined as the first excited state in the physical Hilbert space.

\begin{figure}[t]
 \centering
 \includegraphics[width=1\columnwidth]{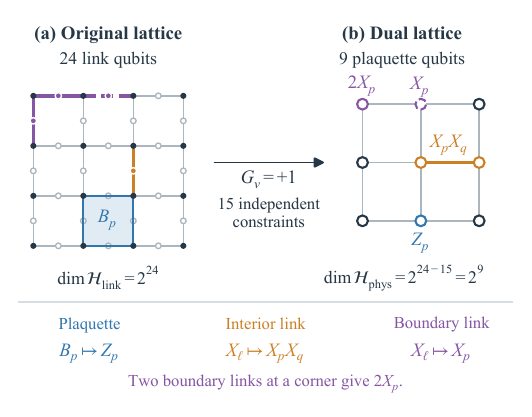}
 \caption{Gauss's law reduction and dual lattice encoding of the $2+1$D $\mathbb{Z}_2$
gauge theory with OBCs. (a) The original $4\times4$ lattice has 24 link qubits. (b) The dual lattice assigns one qubit to each of the nine plaquettes. Imposing the 15 independent constraints $G_v=+1$ reduces the Hilbert space dimension from $2^{24}$ to $2^9$, and the dual representation automatically satisfies Gauss's law. The blue plaquette operator maps as $B_p\mapsto Z_p$, while the orange interior link maps as $X_\ell\mapsto X_pX_q$. A dashed purple boundary link gives a single $X_p$ term.
The two solid purple boundary links are adjacent to the same corner plaquette, so their contributions combine into $2X_p$ in the dual Hamiltonian.}
 \label{fig:Z2}
\end{figure}

To work directly in the physical Hilbert space and simplify state preparation, A2 chooses OBCs and a dual-variable formulation that removes the non-physical degrees of freedom~\cite{Wegner1971Duality,Yamamoto:2020eqi}. The original lattice is shown in Fig.~\ref{fig:Z2}(a), while its dual lattice can be constructed by representing each plaquette as a dual site, which is shown in Fig.~\ref{fig:Z2}(b). A3 maps these dual variables to qubits as $B_p\mapsto Z_p$ and $X_\ell\mapsto\prod_{p\in A(\ell)}X_p$, where $A(\ell)$ contains the plaquettes adjacent to link $\ell$. An interior link produces an $X_pX_q$ term, whereas a boundary link produces a single $X_p$ term. With $a=1$, the resulting dual Hamiltonian is
\begin{equation}
    H_{\rm dual}=-g^{-2}\sum_p Z_p -g^2\left(\sum_{\langle p,q\rangle}X_pX_q+\sum_{p\in\partial P}w_pX_p\right),
 \label{eq:z2_dual_hamiltonian}
\end{equation}
which is defined on a $3\times3$ dual lattice. 
Here $w_p$ counts the number of boundary links belonging to plaquette $p$: $w_p=2$ for the four corners plaquettes and $w_p=1$ for the four non-corner boundary plaquettes. These are indicated by the solid and dotted purple elements in Fig.~\ref{fig:Z2}, respectively. 
Gauss's law is automatically satisfied in the dual representation: for each vertex $v$, every adjacent dual operator $X_p$ appears twice in the product $\prod_{\ell\ni v}X_{\ell}$, so that $G_v \mapsto I$.
The elimination of non-physical degrees of freedom can be verified by counting Hilbert space dimension. The original $4\times4$ lattice has 24 links and 16 vertices. Since one of the 16 Gauss's law constraints is redundant, there are 15 independent constraints, and the physical Hilbert space has dimension $2^{24-15}=2^9$. The original lattice has 9 plaquettes, corresponding to nine qubits on the $3\times3$ dual lattice, which also has dimension $2^9$.

A4.1 chooses a Hamiltonian variational ansatz~\cite{Wiersema2020HVA} built from the generators in Eq.~\eqref{eq:z2_dual_hamiltonian}. Each layer applies nine $R_Z$ rotations, eight boundary $R_X$ rotations, and twelve nearest-neighbor $R_{XX}$ rotations, with 29 independent parameters. The vacuum reference state is $|+\rangle^{\otimes9}$, while the excited reference state replaces the central plaquette by $|-\rangle$. Separate circuits are optimized for the two states, so the orthogonality of the reference states is not automatically preserved. The loss function of the excited state includes an overlap penalty against the optimized vacuum~\cite{Higgott2019VQD}. A5 implements these circuits in Qiskit, and the VQE runtime scans ansatz depths $D=1,2,4,8$, using the optimized parameters at one depth to initialize the next.

\begin{figure}[t]
 \centering
 \includegraphics[width=1\columnwidth]{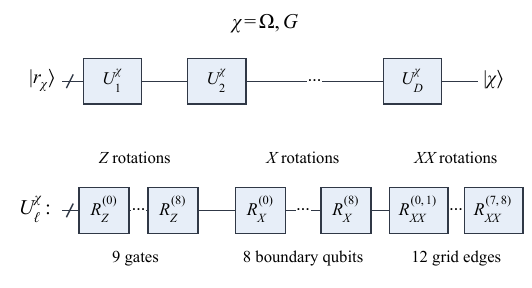}
 \caption{State preparation ansatz for the $2+1$D $\mathbb{Z}_2$ gauge theory. The agent uses different ansatz $V_D^\chi(\theta) = U_D^\chi ... U^\chi_{2} U^\chi_1$, with $\chi = \Omega, G$, to prepare the vacuum and ``glueball'' states from the corresponding reference states $\ket{r_\chi}$. 
The lower circuit shows one layer $U_\ell^\chi$, consisting of nine single qubit $R_Z$ rotations, eight boundary $R_X$ rotations, and twelve nearest neighbor $R_{XX}$ rotations on the dual lattice edges. The superscripts label the qubits on which the gates act. For example, $R_X^{(q)} = \exp(-i \theta_{\ell,q} X_q/2)$ acts on qubit $q$, while $R^{(p,q)}_{XX}$ acts on a nearest neighbor pair $\langle p,q\rangle$. Each layer contains 29 independent parameters.}
 \label{fig:Z2_ansatz}
\end{figure}

For this static observable, A4.2 selects direct measurements of the 29 Pauli terms in each state, with $\Delta E$ reconstructed from their weighted sum. A5 implements the corresponding measurement circuits. For the two states, 58 Pauli expectation values are measured, with the coefficients $w_p$ in Eq.~\eqref{eq:z2_dual_hamiltonian} included when reconstructing the energies. Tab.~\ref{tab:z2_results} compares the noiseless simulation results with ED and TN calculations from the independent programs generated by A6 and A7~\cite{Virtanen2020SciPy,Fishman2022ITensor}. In the numerical cross-check, the ED and TN results differ by $2.42\times10^{-13}$, providing a reference for assessing the variational states.

\begin{table}[t]
 \caption{Energies and gap for the open $4\times4$-vertex $\mathbb{Z}_2$ lattice at $g=1$. The IBM results use $D=1$ and include measurement mitigation and ZNE. The uncertainties are propagated statistical errors.}
 \label{tab:z2_results}
 \begin{ruledtabular}
 \begin{tabular}{@{}lrrr@{}}
 Method & $E_\Omega$ & $E_G$ & $\Delta E$\\
 \hline
 ED & $-25.122839$ & $-17.567359$ & $7.555480$\\
 TN & $-25.122839$ & $-17.567359$ & $7.555480$\\
 NS, $D=1$ & $-25.110287$ & $-17.360130$ & $7.750157$\\
 NS, $D=2$ & $-25.122776$ & $-17.425902$ & $7.696873$\\
 NS, $D=4$ & $-25.122828$ & $-17.516658$ & $7.606171$\\
 NS, $D=8$ & $-25.122834$ & $-17.565535$ & $7.557299$\\
 IBM, $D=1$ & $-25.006\pm0.089$ & $-16.858\pm0.101$ & $8.148\pm0.135$\\
 \end{tabular}
 \end{ruledtabular}
\end{table}

As $D$ increases, the error in $\Delta E$ decreases from $0.19468$ at $D=1$ to $0.001819$ at $D=8$, showing systematic improvement with ansatz depth. At $D=8$, the vacuum energy error is $5.14\times10^{-6}$, whereas the excited-state error remains $1.82\times10^{-3}$. The residual gap error is therefore dominated by the excited state. 

Following the resource assessment by A9, the hardware runtime executes only the $D=1$ demonstration circuits on \texttt{ibm\_kawasaki}. The measured gap, $\Delta E_{\rm hw}=8.1483\pm0.1350$, differs from the matching noiseless value $7.7502$ by approximately $2.95$ standard errors. 
Measurement mitigation and ZNE~\cite{Temme2017ZNE} are applied. 
In the 58 expectation values, 43 use the extrapolated values, while 15 revert to their original values because the extrapolated uncertainties fail the stability criterion. The result demonstrates the hardware execution and reconstruction of the gap at $D=1$, including the propagated hardware uncertainty.

\section{Summary and Outlook}
\label{discussion}

We have developed \texttt{Jianmu-Q}, a multi-agent workflow that turns a concise description of a QFT, its external states, and an observable into executable quantum circuits and a documented numerical analysis. Specialized agents handle continuum formulation, lattice discretization, qubit mapping, and circuit design. Review and execution feedback connect these stages so that errors can be corrected where they arise.

The three benchmarks show that the workflow operates across fermionic and gauge-theory problems in one and two spatial dimensions. In every case, independently generated ED and TN results agree closely. Noiseless simulation results also approach these references as the ansatz depth increases and, for real-time calculations, as the number of Trotter steps increases. These checks demonstrate consistency among the physical definitions, classical implementations, and quantum circuits throughout the automated workflow. Hardware tests further show that the workflow can assess resources, execute selected circuits, and report statistical and mitigation uncertainties. The main significance of \texttt{Jianmu-Q} is therefore methodological. Field-theory simulation normally requires coordinated expertise in continuum theory, lattice methods, many-body numerics, quantum algorithms, and hardware. Assigning these tasks to specialized agents while preserving shared definitions and traceable intermediate results substantially lowers the barrier to multidisciplinary work.

Future work will extend both scale and scope. When exact references become unavailable, validation can combine smaller-lattice comparisons, symmetry tests, controlled limits, state diagnostics, and self-verification~\cite{Kokail2019Verification}. Improvements in state preparation, time evolution, and error mitigation will enable more accurate hardware calculations~\cite{Molpeceres:2026tuw}. Extending \texttt{Jianmu-Q} to thermal states, renormalized observables, and continuum extrapolations will require additional representations and algorithms~\cite{Jordan2012QFT,Bauer:2022hpo,Lu:2021yko,Lu:2024utm}, while retaining the same modular workflow and validation principles.

\section*{Note} \paragraph*{Code availability}
The \texttt{Jianmu-Q} workflow (agents, controller, and numerical runtimes), together with the configurations and saved results of the three benchmarks, including the hardware runs, and the resource-usage records of the supplementary runs in Appendix~\ref{app:resource_usage}, is available at
\url{https://github.com/Jianmu-lab/Jianmu-Q}.
\paragraph*{Use of AI tools} Claude Opus 5.5/Fable 5.1 (Anthropic) and ChatGPT 6.0 (OpenAI) assisted substantially with numerical cross-checks and with polishing the manuscript. All derivations, results, and statements were verified by the authors, who take full responsibility for the content.

\begin{acknowledgments}
We thank Tetsuo Hatsuda, Sirui Lu, Wei Wang, Xiaoyang Wang, and other QuNu collaboration members for insightful discussions. This research was supported by the RIKEN TRIP initiative (RIKEN Quantum).
We thank the DEEP-IN working group at RIKEN-iTHEMS for support in the preparation of this paper.
XW and HX are supported by National Natural
Science Foundation of China with Project No. 12525508, LW is supported by the JST-BOOST Grant No.\ JPMJBY24H9, and JSPS KAKENHI Grant No.\ 25H01560. LW also thanks the support from Google TPU and DeepMind teams.
\end{acknowledgments}

\appendix
\section{Example of an unsuccessful circuit calculation}
\label{app:failed_crosscheck}

Independent reference calculations help distinguish successful program execution from an accurate physical calculation. We illustrate their role with an earlier Schwinger-model calculation, separate from the benchmark in Sec.~\ref{sec:Schwinger}. This calculation used $m=g=0.5$, $a=1$, and five spatial sites, and evaluated
\begin{equation}
 \mathcal{M}(z)=\langle h|\bar\psi(zn)\Gamma W(zn,0)\psi(0)|\Omega\rangle,
 \qquad n=(1,1),
 \label{eq:appendix_failed_target}
\end{equation}
at $z=-2,-1,0,1,2$. Here $|\Omega\rangle$ and $|h\rangle$ are the lowest and first excited states in the neutral sector. The archived input of this earlier run specified $\Gamma=\gamma^0+\gamma^1$ and the bra--ket ordering shown in Eq.~\eqref{eq:appendix_failed_target}. These choices differ from the target in Sec.~\ref{sec:Schwinger}; in particular, $\Gamma$ here is specified independently of $n$ and is not identified with $n\cdot\gamma=\gamma^0-\gamma^1$. This run is therefore treated as a separate diagnostic example.

Since a transition matrix element depends on the arbitrary relative phase of the external eigenstates, we align this phase before comparing different methods, as in the main Schwinger benchmark. For each ED, TN, and noiseless-simulation (NS) curve, we rotate $\mathcal{M}(z)$ in the complex plane by a constant phase such that $\operatorname{Im}\mathcal{M}(0)=0$ and $\operatorname{Re}\mathcal{M}(0)>0$. The same rotation is applied to all five values of $z$ within each curve. The NS series is evaluated at the deepest available setting, $(D,N_T)=(4,16)$.

\begin{table*}[t]
 \caption{Phase-aligned transition matrix elements for the earlier Schwinger-model run. NS denotes noiseless circuit simulation at $(D,N_T)=(4,16)$. Values are rounded for display; the discrepancies are computed from the unrounded archived data. ED and TN give closely matching values, whereas NS shows substantial deviations at nonzero separations.}
 \label{tab:failed_schwinger_crosscheck}
 \begin{ruledtabular}
 \begin{tabular}{rrrrr}
 $z$ & $\mathcal{M}_{\rm ED}$ & $\mathcal{M}_{\rm TN}$ & $\mathcal{M}_{\rm NS}$ & $|\mathcal{M}_{\rm NS}-\mathcal{M}_{\rm ED}|$\\
 \hline
 $-2$ & $0.177550-0.194599i$ & $0.177551-0.194599i$ & $-0.128967+0.162527i$ & $0.470629$\\
 $-1$ & $0.277118-0.147347i$ & $0.277117-0.147347i$ & $0.110720+0.219512i$ & $0.402832$\\
 $0$ & $0.301349$ & $0.301349$ & $0.305577$ & $0.004229$\\
 $1$ & $0.177421+0.098362i$ & $0.177404+0.098487i$ & $0.155953-0.201501i$ & $0.300630$\\
 $2$ & $0.072926+0.041414i$ & $0.072513+0.040118i$ & $-0.006246-0.238848i$ & $0.291230$\\
 \end{tabular}
 \end{ruledtabular}
\end{table*}

Table~\ref{tab:failed_schwinger_crosscheck} shows substantial NS--ED discrepancies even after phase alignment. The largest difference is $0.471$, whereas the maximum ED--TN difference after alignment is $1.36\times10^{-3}$. At $z=0$, the NS value differs from ED by only $0.00423$, but agreement at this point does not extend to the nonzero separations. Thus the disagreement cannot be explained solely by an arbitrary constant phase. These diagnostics identify an unsuccessful observable calculation, although they do not by themselves determine whether its dominant error comes from state preparation, measurement implementation, or finite-step evolution.

The close agreement between ED and TN provides a consistent numerical reference for this comparison. Their maximum difference is more than two orders of magnitude smaller than the largest NS discrepancy. This agreement supports identifying the NS calculation as the source of the substantial deviation, rather than a disagreement between the two classical methods. This case illustrates why a generated circuit that executes and returns a smooth set of values still needs observable-level numerical validation. Without the reference comparisons, this direct evidence of the discrepancy would be absent; energy diagnostics and internal consistency checks alone would not establish the accuracy of the requested matrix element.

\section{User input and computational resource usage}
\label{app:resource_usage}

To quantify the work performed from a concise scientific specification, we report supplementary resource measurements for tasks in the NJL, Schwinger, and pure $\mathbb Z_2$ model classes. These measurements use GPT-6 Sol (medium reasoning effort), whereas the main benchmarks were performed with GPT-5.6 Sol.
The supplementary calculations were carried out to record token consumption and LLM-call duration; they do not replace the numerical results presented in the main text. Since they follow the same staged workflow for the same classes of physical tasks, their token counts provide an indicative order-of-magnitude reference for the LLM cost of the workflow. Exact counts can vary with the model, generated algorithms, and retries; a paired comparison of token usage between the two models was not recorded.

The measurements are taken from the resource-usage records of the three supplementary calculations, including recorded failures and retries.

\paragraph{User specification and internal context.}
The recorded user-input lengths are local estimates using the \texttt{o200k\_base} encoding. Each counts the submitted scientific fields once, serialized as compact sorted JSON, including field names, punctuation, and UI selections or defaults. Clarification answers are counted separately. These lengths exclude system prompts, retrieved literature, generated intermediate results, and context passed repeatedly between agents; they are not API billing counts.

\begin{table*}[ht]
 \caption{User-input estimates and cumulative LLM usage in the supplementary GPT-6 Sol (medium reasoning effort) runs. User and clarification tokens are local estimates; LLM input and output tokens are reported API usage, including recorded retries and the A10 call. LLM time is the cumulative duration of recorded calls, including network waiting and generation.}
 \label{tab:llm_resource_totals}
 \begin{ruledtabular}
 \small
 \begin{tabular}{lrrrrrrr}
 Task & User tokens & Clarification & LLM calls & LLM input & LLM output & LLM total & LLM time\\
 & & tokens & & tokens & tokens & tokens & (min)\\
 \hline
 NJL & 374 & 0 & 21 & 578,614 & 131,530 & 710,144 & 18.32\\
 Schwinger & 316 & 0 & 17 & 405,844 & 98,595 & 504,439 & 15.50\\
 $\mathbb Z_2$ & 181 & 16 & 16 & 239,284 & 62,615 & 301,899 & 10.00\\
 \end{tabular}
 \end{ruledtabular}
\end{table*}

Table~\ref{tab:llm_resource_totals} highlights the concise user interface: the three tasks require only approximately 181--374 tokens of original scientific input, with 0--16 additional clarification tokens, yet initiate 16--21 LLM calls and 62,615--131,530 output tokens. For example, the NJL task starts from approximately 374 user-input tokens and produces 131,530 output tokens across its agents. The user specifies the physical task, while the workflow supplies the stage-specific instructions and coordinates derivations, code generation, review, and numerical execution. This supports the practical accessibility of the interface: users can initiate a multidisciplinary calculation without composing a separate detailed prompt for every stage. The larger API input totals additionally include internal prompts, retrieved material, and intermediate results passed between agents; these repeated contexts are distinct from newly generated output.

\paragraph{Agent-level accounting.}
Table~\ref{tab:llm_resource_agents} resolves usage by agent. A5 has separate state-preparation and measurement calls. Each entry accumulates the recorded calls associated with that stage, including scientific review and retries where present. Input tokens include context supplied to the model; output tokens include the reported reasoning tokens. Reasoning tokens are therefore already included in the totals and are not counted separately. Their recorded totals are 38,170, 32,689, and 16,988 for NJL, Schwinger, and $\mathbb Z_2$, respectively. All three records report zero cached-input tokens, and every recorded call includes usage information.

\begin{table*}[thbp]
 \caption{Recorded agent-level resource usage with GPT-6 Sol (medium reasoning effort). Tokens are the sum of API input and output tokens, and LLM time includes network waiting and generation. Entries include recorded retries; each column sums to the corresponding total in Table~\ref{tab:llm_resource_totals}, up to rounding of the time entries.}
 \label{tab:llm_resource_agents}
 \begin{ruledtabular}
 \small
 \begin{tabular}{lrrrrrr}
 & \multicolumn{2}{c}{NJL} & \multicolumn{2}{c}{Schwinger} & \multicolumn{2}{c}{$\mathbb Z_2$}\\
 Agent & Tokens & LLM (s) & Tokens & LLM (s) & Tokens & LLM (s)\\
 \hline
A1 & 24,158 & 111.46 & 17,020 & 84.16 & 12,258 & 56.46 \\
A2 & 39,343 & 110.91 & 44,960 & 149.70 & 34,927 & 88.50 \\
A3 & 25,369 & 81.58 & 32,376 & 109.15 & 22,243 & 68.85 \\
A4.1 & 63,316 & 168.72 & 28,423 & 59.89 & 25,333 & 58.17 \\
A4.2 & 31,299 & 101.62 & 31,820 & 96.99 & 27,333 & 62.28 \\
A5 (state) & 42,844 & 104.68 & 20,564 & 49.89 & 17,282 & 44.34 \\
A5 (measurement) & 101,258 & 198.32 & 23,225 & 66.72 & 21,155 & 57.13 \\
A6 & 11,860 & 49.82 & 35,004 & 116.08 & 10,627 & 45.42 \\
A7 & 15,310 & 71.26 & 17,742 & 90.03 & 12,582 & 56.36 \\
A8 & 40,961 & 28.37 & 30,897 & 28.50 & 15,966 & 10.45 \\
A9 & 21,053 & 26.16 & 21,305 & 21.35 & 2,622 & 14.02 \\
A10 & 293,373 & 46.54 & 201,103 & 57.64 & 99,571 & 37.75 \\
 \end{tabular}
 \end{ruledtabular}
\end{table*}

\clearpage

\bibliography{refs}

\end{document}